\documentclass[12pt]{article}
\usepackage{a4wide,amssymb,float}
\usepackage{latexsym,epsfig,graphicx}

\def\be{\begin{equation}}
 \def\ee{\end{equation}}
 \def\bea{\begin{eqnarray}}
 \def\eea{\end{eqnarray}}
\newcommand{\fr}{\frac}
\newcommand{\pr}{\prime}

\def\2{\frac{1}{2}}
\def\4{\frac{1}{4}}

\catcode`\@=11

\@addtoreset{equation}{section}

\def\@normalsize{\@setsize\normalsize{15pt}\xiipt\@xiipt
\abovedisplayskip 14pt plus3pt minus3pt%
\belowdisplayskip \abovedisplayskip
\abovedisplayshortskip  \z@ plus3pt%
\belowdisplayshortskip  7pt plus3.5pt minus0pt}
\def\small{\@setsize\small{13.6pt}\xipt\@xipt
\abovedisplayskip 13pt plus3pt minus3pt%
\belowdisplayskip \abovedisplayskip
\abovedisplayshortskip  \z@ plus3pt%
\belowdisplayshortskip  7pt plus3.5pt minus0pt
\def\@listi{\parsep 4.5pt plus 2pt minus 1pt
            \itemsep \parsep
            \topsep 9pt plus 3pt minus 3pt}}

\def\underline#1{\relax\ifmmode\@@underline#1\else
        $\@@underline{\hbox{#1}}$\relax\fi}
\@twosidetrue \relax

\catcode`@=12

\catcode`\@=11

\def\section{\@startsection{section}{1}{\z@}{3.5ex plus 1ex minus
   .2ex}{2.3ex plus .2ex}{\large\bf}}
\def\ps@headings{\def\@oddfoot{}\def\@evenfoot{}
\def\@oddhead{\hbox{}\hfill
        \makebox[.5\textwidth]{\raggedright\ignorespaces --\thepage{}--
        \hfill }}
\def\@evenhead{\@oddhead}
\def\subsectionmark##1{\markboth{##1}{}}
}

\ps@headings

\catcode`\@=12

\begin{document}

\begin{titlepage}
\begin{flushright}
hep-th/0606096\\ May 2006
\end{flushright}

\vspace{0.30in}
\begin{centering}

{\large {\bf Quasi-normal Modes of Electromagnetic Perturbations of
Four-Dimensional Topological Black Holes with Scalar Hair}}
\\

\vspace{0.7in} {\bf George Koutsoumbas,$^{a,*}$\,\,\,\, Suphot
Musiri,$^{b,\dag}$\,\,\,\,Eleftherios
Papantonopoulos$^{a,\flat}$\,\,\,\,\\ \vspace{0.14in} and George
Siopsis$^{c,\natural}$}

\vspace{0.44in}

$^{a}$Department of Physics, National Technical University of Athens,\\
Zografou Campus GR 157 73, Athens, Greece\\
$^{b}$Department of Physics, Srinakharinwirot University, Bangkok
10110,
Thailand\\
$^{c}$Department of Physics and Astronomy, The University of
Tennessee, Knoxville, TN 37996 - 1200, USA

\vspace{0.04in}

\vspace{0.04in}

\end{centering}

\vspace{0.8in}

\begin{abstract}

We study the perturbative behaviour of topological black holes
with scalar hair. We calculate both analytically and numerically
the quasi-normal modes of the electromagnetic perturbations. In
the case of small black holes we find evidence of a second-order
phase transition
%in the vicinity of a critical temperature
of a
topological black hole to a hairy configuration.
%We also find
%evidence of a second-order phase transition of the AdS vacuum
%solution to a topological black hole.

 \end{abstract}

\begin{flushleft}

\vspace{0.9in} $^{*}$~kutsubas@central.ntua.gr \\
$^\dag$~suphot@swu.ac.th\\
$^{\flat}$~lpapa@central.ntua.gr\\
$^\natural$~siopsis@tennessee.edu

\end{flushleft}
\end{titlepage}

\section{Introduction}
%%%%%%%%%%%%%%%%%%%%%%%%%%%%%MTZ%%%%%%%%%%%%%%%%%%%%%%%%%%%%

Quasi-normal modes (QNMs) are well known to play an important role
in black hole physics. They determine the late-time evolution of
fields in the black hole exterior. After an initial perturbation
the black hole starts vibrating into quasi-normal oscillation
modes whose frequencies and decay times depend only on the
intrinsic features of the black hole itself being insensitive to
the details of the initial perturbation. For these reasons, QNMs
of black holes in asymptotically flat spacetimes have been
extensively studied (for reviews, see~\cite{KS,N}).

The Anti-de Sitter - conformal field theory (AdS/CFT)
correspondence has led to an intensive investigation of black hole
QNMs in asymptotically AdS spacetimes. Quasi-normal modes in AdS
spacetime were first computed for a conformally invariant scalar
field, whose asymptotic behaviour is similar to flat
spacetime~\cite{CM}.  Subsequently, motivated by the AdS/CFT
correspondence, Horowitz and Hubeny made a systematic computation
of QNMs for scalar perturbations of Schwarzschild-AdS (S-AdS)
spacetimes~\cite{HH}.  Their work was extended to gravitational
and electromagnetic perturbations of S-AdS black holes
in~\cite{CL}. The study of scalar perturbations was further
extended to the case of Reissner-Nordstr\"om-AdS (RN-AdS) black
holes in~\cite{WLA}. Finally, the QNMs of  scalar, electromagnetic
and gravitational perturbations of RN-AdS black holes were
presented in~\cite{kokkotas} using the results of~\cite{MM}.

In a parallel development, exact black hole solutions in
asymptotically non-flat spacetimes were studied recently and
solutions with scalar hair and negative cosmological constant were
found. Exact black hole solutions are known in three~\cite
{Martinez:1996gn,Henneaux:2002wm} and four
dimensions~\cite{Martinez:2002ru}. For asymptotically flat
spacetime, a four-dimensional black hole is also known, but the
scalar field diverges at the horizon~\cite{BBMB}. Also spherically
symmetric black hole solutions have been found numerically in
four~\cite{Torii:2001pg,Winstanley:2002jt} and five
dimensions~\cite{Hertog:2004dr}. Spaces with a negative
cosmological constant also allow for the existence of black holes
whose horizon has nontrivial topology in four~\cite{Lemos,mann,
Vanzo:1997gw,Brill:1997mf} and higher
dimensions~\cite{Birmingham,Cai-Soh,Wang:2001tk,MannTHB} as well
as for gravity theories containing higher powers of the
curvature~\cite {BHscan,Aros:2000ij,Cai-GB,Dehghani}.

Recently, an exact black hole solution in four dimensions with a
minimally coupled self-interacting scalar field, in an
asymptotically locally anti-de Sitter spacetime, was
found~\cite{Martinez:2004nb} (MTZ black hole). The event horizon is a
surface of negative constant curvature enclosing the curvature
singularity at the origin. It was shown that there is a second-order
 phase transition at a critical temperature  below which a
black hole in vacuum undergoes a spontaneous dressing up with a
nontrivial scalar field. An extension of the above solution
including a charge was presented in~\cite{Martinez:2005di}.
Aspects of the thermodynamics of the MTZ black hole are discussed
in~\cite{winstanley}. In~\cite{papadimitriou} it was shown that
the four-dimensional MTZ black hole can be uplifted to eleven
dimensions in supergravity theory.

In this work, we make a perturbative study of the MTZ black hole.
We compute the simplest possible QNMs of the MTZ black hole, those
of electromagnetic (EM) perturbations. The computation is carried
out both analytically and numerically with fairly good agreement.
The QNMs provide, near a critical temperature and for small black
holes, support for the claim that a vacuum topological black hole
(TBH) goes over to a hairy configuration, the MTZ black hole,
through a second-order phase transition.

The paper is organized as follows. In section~\ref{sec2}, we
review the MTZ black hole and its charged extension. In
section~\ref{sec3}, we discuss the thermodynamics of the MTZ and
TBH black holes. In section~\ref{sec4}, we calculate the QNMs of
the EM perturbations analytically whereas in Section~\ref{sec5} we
present the numerical computation of the QNMs and compare them
with the analytical results of section~\ref{sec4}. Finally,
section~\ref{sec6} contains our conclusions.

\section{Four-Dimensional Topological Black Hole with Scalar Hair}
\label{sec2}

Consider four-dimensional gravity with negative cosmological constant ($%
\Lambda =-3l^{-2}$) and a scalar field described by the action
\begin{equation}
I=\int d^{4}x\sqrt{-g}\left[ \frac{R+6l^{-2}}{16\pi G}-%
\frac{1}{2}g^{\mu \nu }\partial _{\mu }\phi \partial _{\nu }\phi
-V(\phi )\right] \;,  \label{action}
\end{equation}
where $l$ is the AdS radius, and $G$ is the Newton's constant. The
 self-interaction potential is given by
\begin{equation}
V(\phi )=-\frac{3}{4\pi Gl^{2}}\sinh ^{2}\sqrt{\frac{4\pi
G}{3}}\phi \;, \label{potential}
\end{equation}
which has a global maximum at $\phi =0$, and has a mass term given by $%
m^{2}=\left. V^{\prime \prime }\right| _{\phi =0}=-2l^{-2}$. This
mass satisfies the Breitenlohner-Friedman bound that ensures the
perturbative stability of AdS spacetime~\cite{B-F,M-T}.
  The field equations are
\begin{eqnarray}
G_{\mu \nu }-\frac{3}{l^{2}}g_{\mu \nu } &=&8\pi G\,T_{\mu \nu
}\;,
\nonumber \\
\square \phi -\frac{dV}{d\phi } &=&0\;,  \label{Feq}
\end{eqnarray}
where $\square \equiv g^{\mu \nu }\nabla _{\mu }\nabla _{\nu }$,
and the stress-energy tensor is given by
\begin{equation}
T_{\mu \nu }=\partial _{\mu }\phi \partial _{\nu }\phi
-\frac{1}{2}g_{\mu \nu }g^{\alpha \beta }\partial _{\alpha }\phi
\partial _{\beta }\phi -g_{\mu \nu }V(\phi )\;.  \label{Tuv}
\end{equation}
A static black hole solution with topology $\mathbb{R}^{2}\times
\Sigma $, where $\Sigma $ is a two-dimensional manifold of
negative constant curvature, is given by~\cite{Martinez:2004nb}
\begin{equation}
ds^{2}=\frac{r(r+2G\mu )}{(r+G\mu )^{2}}\left[ -\left( \frac{r^{2}}{l^{2}}%
-\left( 1+\frac{G\mu }{r}\right) ^{2}\right) dt^{2}+\left( \frac{r^{2}}{l^{2}%
}-\left( 1+\frac{G\mu }{r}\right) ^{2}\right)
^{-1}dr^{2}+r^{2}d\sigma ^{2}\right],  \label{Black-Hole}
\end{equation}
and the scalar field is
\begin{equation}
\phi =\sqrt{\frac{3}{4\pi G}}\;\mbox{Arctanh}\frac{G\mu }{r+G\mu
}\;. \label{scalarmtz}
\end{equation}  The
position of the horizon is at $r_{+}$, which is the solution of
  \be G\mu=\frac{r^{2}_{+}}{l}-r_{+}~.\label{gm}
\ee For $\phi=0$ we get the vacuum solution (Topological Black
Hole (TBH))~\cite{Lemos,mann, Vanzo:1997gw,Brill:1997mf}
\begin{equation}
ds_{0}^{2}=-\left[ \frac{r ^{2}}{l^{2}}-1-\frac{2G\mu }{r }\right]
dt^{2}+\left[ \frac{r ^{2}}{l^{2}}-1-\frac{2G\mu }{r }\right]
^{-1}dr ^{2}+r ^{2}d\sigma ^{2}\;. \label{phizero}\ee

A charged black hole with scalar hair was presented
in~\cite{Martinez:2005di}. The action is given by
\begin{eqnarray}
I&=&\int d^4x\sqrt{-g}\left[\frac{R+6l^{-2}}{16\pi
G}-\frac{1}{2}g^{\mu\nu}\partial_\mu\phi\partial_\nu\phi\right.
-\left.\frac{1}{12}R\phi^2-\alpha\phi^4\right] \nonumber\\
\label{action1}& &-\frac{1}{16\pi }\int d^{4}x\sqrt{-g}F^{\mu \nu
}F_{\mu \nu }~,
\end{eqnarray}
where $\alpha$ is an arbitrary coupling constant.

The corresponding field equations are
\begin{eqnarray}
G_{\mu\nu} -\frac{3}{l^{2}} g_{\mu\nu}&=&8\pi G (T_{\mu\nu}^{\phi}+T_{\mu\nu}^{\rm{em}})~, \nonumber\\
\square\phi&=&\frac{1}{6}R\phi+\alpha\phi^3~, \nonumber\\
\partial _{\nu }(\sqrt{-g}F^{\mu \nu }) &=&0~,
\end{eqnarray}
%where $\square \equiv g^{\mu \nu }\nabla _{\mu }\nabla _{\nu }$,
and the energy-momentum tensor is given by the sum of
\begin{equation}
\label{Tuvfield}T_{\mu\nu}^{\phi}=\partial_\mu\phi\partial_\nu\phi-\frac{1}{2}
g_{\mu\nu}g^{\alpha\beta}\partial_\alpha\phi\partial_\beta\phi
 +\frac{1}{6}[g_{\mu\nu}\square-\nabla_\mu\nabla_\nu+G_{\mu\nu}]\phi^2
-g_{\mu\nu}\alpha\phi^4,
\end{equation}
and
\begin{equation}
\label{Tuvem}T^{\rm{em}}_{\mu \nu }=\frac{1}{4\pi } \left( F_{\mu
\alpha}F_{\nu \beta} -\frac{1}{4}g_{\mu \nu } F_{\gamma
\alpha}F_{\delta \beta} g^{\gamma \delta } \right) g^{\alpha \beta
}.
\end{equation}
%Note that Eq. (\ref{kg}) can be obtained from Eq. (\ref{ee}) using
%the conservation of the energy-momentum tensor. Since the scalar
%field is conformally coupled, $T_{\mu\nu}^{\phi}$ is traceless as
%well as $T^{\rm{em}}_{\mu \nu }$. This traceless property yields a
%simple expression for Ricci scalar curvature $R=4\Lambda$
The charged static black hole solution with topology
$\mathbb{R}^{2}\times \Sigma $ is given by
\begin{equation} \label{bhsolution}
\label{solution}ds^2=-\Big[ \frac{r
^{2}}{l^{2}}-\Big(1+\frac{G\mu}{r}\Big)^2\Big]dt^2+\Big[ \frac{r
^{2}}{l^{2}}-\Big(1+\frac{G\mu}{r}\Big)^2\Big]^{-1}dr^2+r^2
d\sigma^2,
\end{equation}
where  $-\infty < t < \infty$ and $r>0$. The scalar field is
\begin{equation} \label{scalar3}
\phi=\sqrt\frac{1}{2\alpha l^{2}}\frac{G\mu}{(r+G\mu)}~,
\end{equation}
with $\alpha > 0$ and the only non-zero component of the
electromagnetic field is
\begin{equation} \label{electric}
A_{t}=-\frac{q}{r}\;.
\end{equation}
The integration constants $q$ and $\mu$ are not independent. They
are related by
\begin{equation} \label{ratio1}
q^{2}= - G\mu ^{2}\left( 1-\frac{2\pi  G}{3\alpha l^{2}}\right)
\;.
\end{equation}
They correspond to conserved
charges,
\begin{equation}
M=\frac{\sigma }{4\pi }\mu \quad \mbox{and} \quad
Q=\frac{\sigma}{4\pi }q~,  \label{BlackHoleMass}
\end{equation}
respectively, where $\sigma $ denotes the area of $\Sigma $.

Eq.~(\ref{ratio1}) fixes a charge-to-mass ratio for this
black hole, which is a function of the constants appearing in the
action, $G$ and $\alpha$. Moreover, eq.~(\ref{ratio1}) determines an
upper bound for $\alpha$
\begin{equation}
0 < \alpha \leq \frac{2\pi G}{3l^{2}}~.
\end{equation}
If the upper bound is saturated, the charge vanishes. Then we
recover the MTZ black hole. Indeed the form of the
self-interacting potential considered in (\ref{potential}) can be
naturally obtained through the relation between the conformal and
Einstein frames~\cite{Martinez:2004nb}.

Note that if $\mu=0$ then both the MTZ  black hole
(\ref{Black-Hole}) and the TBH black hole (\ref{phizero}) go to
\begin{equation}
ds_{AdS}^{2}=-\left[ \frac{r ^{2}}{l^{2}}-1\right] dt^{2}+\left[
\frac{r ^{2}}{l^{2}}-1\right] ^{-1}dr ^{2}+r ^{2}d\sigma ^{2}\;,
\label{muzero}\ee which is a manifold of negative constant
curvature possessing an event horizon at $r=l$. We can say that as
$ \phi \rightarrow 0 $ the MTZ and TBH black holes match
continuously at $\mu=0$ or $ r=l$ with (\ref{muzero}) being a
transient configuration as it becomes apparent in the following.
In the sequel we set $l=1.$

\section{Thermodynamics}
\label{sec3}

For the MTZ black hole, the temperature, entropy and  mass are
given respectively by~\cite{Martinez:2004nb} \bea T&=&\frac{2r_{+}-1}{2 \pi}~,
\nonumber
\\S&=&\frac{\sigma r_{+}(r_{+}+2G\mu )}{4G(r_{+}+G\mu
)^{2}}r^{2}_{+}=\frac{\sigma}{4G}(2r_{+}-1)~,\nonumber \\
M&=&\frac{\sigma \mu}{4 \pi}=\frac{\sigma(r^{2}_{+}-r_{+})}{4 \pi
G}~.\label{relations1} \eea Notice that the entropy in this case
does not satisfy an area law. It is easy to show that the law of
thermodynamics $dM=TdS$ holds. Defining the free energy as
$F=M-TS$ and using relations (\ref{relations1}), we obtain \be
F_{MTZ}=-\frac{\sigma}{8\pi G}(2r^{2}_{+}-2r_{+}+1)~.
\label{free-energy}\ee The free energy~(\ref{free-energy}) can be
written as \be F_{MTZ}=-\frac{\sigma}{8\pi
G}\Big{(}1+2(T-T_{0})\pi +2 (T-T_{0})^{2}\pi^{2}
\Big{)}~,\label{mtzexpan}\ee where $T_{0}=1/2\pi \approx 0.160$ is
the critical temperature. For the vacuum TBH black hole (denoting
by $\rho_+$ the horizon for this case) , the temperature, entropy
and mass are, respectively, \bea T&=&\frac{3\rho^{2}_{+}-1}{4 \pi
\rho_{+}}~, \nonumber
\\S&=&\frac{\sigma \rho^{2}_{+}}{4G}~,\nonumber \\
M&=&\frac{\sigma(\rho^{3}_{+}-\rho_{+})}{8 \pi
G}~.\label{relations2} \eea Then, the free energy of the TBH black
hole, using relations (\ref{relations2}), is \be
F_{TBH}=-\frac{\sigma}{16\pi G}(\rho^{3}_{+}+\rho_{+})~,
\label{FETBH} \ee which can be expanded around the critical
temperature $T_{0}$ as \be F_{TBH}=-\frac{\sigma}{8\pi
G}\Big{(}1+2(T-T_{0})\pi +2 (T-T_{0})^{2}\pi^{2}
+(T-T_{0})^{3}\pi^{3}+\dots \Big{)}~.\label{tbhexpan}\ee Using
(\ref{mtzexpan}) and (\ref{tbhexpan}), we can calculate the
difference between the  TBH and MTZ  free energies. We obtain \be
\Delta F= F_{TBH}-F_{MTZ}=-\frac{\sigma}{8\pi
G}(T-T_{0})^{3}\pi^{3}+\dots~,\label{difference} \ee indicating a
phase transition between MTZ and TBH at the critical temperature
$T_0$. Matching the temperatures of the MTZ black hole and the TBH
we get: \be T = \frac{2 r_+-1}{2 \pi} = \frac{1}{4 \pi} \left(3
\rho_+-\frac{1}{\rho_+}\right) \Rightarrow  r_+ = \frac{3
\rho_+}{4}-\frac{1}{4 \rho_+}+\frac{1}{2}~.\label{rhor} \ee It is
easily seen that $r_+ \le \rho_+,$ and the inequality is saturated
for $r_+ = \rho_+=1.$ We remark that the temperature $T$ should be
non-negative, so $r_+ \ge \fr{1}{2}$ for the MTZ black hole and
$\rho_+ \ge \fr{1}{\sqrt{3}}$ for the TBH black hole.

Thermodynamically we can understand this phase transition as
follows. Using relations (\ref{relations1}), (\ref{relations2})
and (\ref{rhor}), we find that $S_{TBH}>S_{MTZ}$ and
$M_{TBH}>M_{MTZ}$ for the relevant ranges of the horizons $r_+$ or
$\rho_+.$ If $r_{+}>1$~ ($T>T_{0}$), both black holes have
positive mass. As $T>T_0$ implies $F_{TBH} \le F_{MTZ},$ the MTZ
black hole dressed with the scalar field will decay into the bare
black hole. In the decay process, the scalar black hole absorbs
energy from the thermal bath, increasing its horizon radius (from
$r_+$ to $\rho_+>r_+$) and consequently its entropy. Therefore, in
a sense the scalar field is absorbed by the black hole.

If $r_{+}<1$~ ($T<T_{0}$), both black holes have negative mass,
but now $F_{TBH} > F_{MTZ},$ which means that the MTZ
configuration with nonzero scalar field is favorable.  As a
consequence, below the critical temperature, the bare black hole
undergoes a spontaneous ``dressing up'' with the scalar field. In
the process, the mass and entropy of the black hole decrease and
the differences in energy and entropy are transferred to the heat
bath.

At the critical temperature, the thermodynamic functions of the
two phases match continuously, hence, the phase transition is of
second order. The order parameter that characterizes the
transition can be defined in terms of the value of the scalar
field at the horizon; using the solution for the scalar field
(\ref{scalarmtz}) we obtain for $T<T_0$, \be\lambda_{\phi} =\left|
\tanh \sqrt{\frac{4\pi G}{3}}\phi (r_{+})\right|~=
\left|\frac{r_{+}-1}{r_{+}}\right| = \frac{T_{0}-T}{T_{0}+T} \ \ .
\ee For $T>T_0,$ $\lambda_\phi$ vanishes. Then the relation
(\ref{difference}) can be written in terms of the order parameter
$\lambda_{\phi}$ as \be \Delta F= F_{TBH}-F_{MTZ}= + \frac{\sigma
r_+^3}{8\pi G}\lambda^{3}_{\phi}+\dots \ \ \ \ (T<T_{0}).
\label{difference1} \ee

The pure AdS space of (\ref{muzero}) has free energy
$F_{AdS}=-\sigma/8\pi G$ as easily can be seen using relations
(\ref{free-energy}) or (\ref{FETBH}) with $r_{+}=1$. Then observe
that $F_{AdS}$ is the constant term of both $F_{MTZ}$ in
(\ref{mtzexpan}) and $F_{TBH}$ in (\ref{tbhexpan}). Hence  the
difference of free energies of MTZ or TBH black holes  with the
free energy of pure AdS space indicates that the configuration
(\ref{muzero}) is transient between the MTZ and TBH phase
transition.

 In the next two sections we will find the QNMs of the
electromagnetic perturbations of the MTZ black hole and its
charged generalization and compare them with the corresponding
QNMs of the TBH. This study will provide additional information on
the stability of the MTZ black hole and its generalization under
electromagnetic perturbations. Note that for the MTZ black hole
and its charged generalization, the wave equations, after
factoring out the angular parts, are the same. For the TBH, only
the function $f(r)$ (eq.~(\ref{eqZ})) changes. There is no change
due to the axial or polar character of the perturbation.

\section{Analytical Calculation}
\label{sec4}

In this section, we calculate analytically the QNMs of
electromagnetic perturbations for both MTZ and topological black holes.
Numerical results will be discussed in section~\ref{sec5}.
Our approach
is based on the method discussed in~\cite{biba,bibb,bibc}.
In order to calculate QNMs, in general, one solves the wave equation subject to
appropriate boundary conditions at infinity and the horizon.
In asymptotically flat spacetime, this can be implemented by a monodromy method~\cite{bib1}.
It is then advantageous to follow Stokes lines which go through the black hole
singularity.
In asymptotically (A)dS spacetimes, the calculation simplifies because the
wavefunction vanishes at infinity.
Although the monodromy does not enter the calculation, in several cases it is still advantageous to follow Stokes lines and analytically continue the wavefunction through the black hole singularity~\cite{biba,bibb,bibc}.
However, this is not always necessary, if a general solution (or approximation thereof) of the wave equation as in, e.g.,~\cite{bib2} is readily available.
In our case, we shall solve the wave equation near the black hole singularity
perturbatively and then analytically continue the wavefunction in order to
match with the expected behaviour at infinity and the horizon, respectively.
This will produce an explicit form of the QNMs for $r_+>1$ ($G\mu >0$).
For $r_+<1$ ($G\mu <0$), we do not have explicit analytical expressions. Instead, we obtain a lower bound which is in accord with numerical results.

Electromagnetic perturbations obey the wave equation
\be\label{eqwav0} f(r)
\frac{d}{dr} \left( f(r) \frac{d\Psi_\omega}{dr} \right) + \left[\omega^2
- \left( \xi^2 + \frac{1}{4}\right) \, \frac{f(r)}{r^2} \right]
\Psi_\omega = 0~, \ee where \be f_{MTZ} (r) = r^2 - \left( 1+
\frac{G\mu}{r} \right)^2 \ \ , \ \ \ \ f_{TBH} (r) = r^2 - 1 -
\frac{2G\mu}{r}~. \label{eqZ} \ee It may be cast into a
Schr\"odinger-like form if written in terms of the tortoise
coordinate defined by \be\label{eqtor} \frac{dx}{dr} =
\frac{1}{f(r)}~. \ee We obtain \be\label{eqwav} -
\frac{d^2\Psi_\omega}{dx^2} + V(x) \Psi_\omega = \omega^2 \Psi_\omega~,\label{schereq}
\ee where the potential is \be V[x(r)] = \left( \xi^2 +
\frac{1}{4}\right) \, \frac{f(r)}{r^2}~. \ee For QNMs, we impose
the boundary condition $\Psi_\omega \to 0$ as $r\to\infty$, since the
potential does not vanish for large $r$. At the horizon ($x\to
-\infty$),  we demand $\Psi_\omega \sim e^{i\omega x}$ (ingoing wave).

Near the black hole singularity ($r\sim 0$), the tortoise
coordinate~(\ref{eqtor}) may be approximated by \be\label{eqtor2}
x \approx -\frac{G\mu}{a\lambda} \left( \frac{r}{G\mu}
\right)^\lambda~,\ee where $a = 1$, $\lambda = 3$ for MTZ and
$a=2$, $\lambda = 2$ for TBH. In arriving at~(\ref{eqtor2}), we
choose the integration constant so that $x=0$ at $r=0$. The
potential near the singularity is \be\label{eqpot} V(x) \approx
-\frac{\mathcal{A}}{x^{1+1/\lambda}} \ \ , \ \ \ \ \mathcal{A} =
\left( \xi^2 + \frac{1}{4}\right) \,
\frac{1}{(aG\mu)^{1-1/\lambda} (-\lambda)^{1+1/\lambda}}~.\ee We
will solve the wave eq.~(\ref{eqwav}) by treating the
potential~(\ref{eqpot}) as a perturbation. The zeroth-order
solutions are \be\label{eqpsi0} \Psi_0^\pm = e^{\pm i\omega x} \ee
and the first-order corrections are \be\label{eqpsi1} \Psi_1^\pm
(x) = \frac{1}{2i\omega} \int_0^x dx' e^{i(\omega -i\epsilon)
(x-x')} V(x') \Psi_0^\pm (x') - \frac{1}{2i\omega} \int_0^x dx'
e^{-i(\omega +i\epsilon) (x-x')} V(x') \Psi_0^\pm (x') \ee where
we included a small $\epsilon >0$ to render integrals finite. We
shall work with general values of $\lambda$ in the potential
(\ref{eqpot}) and take the limit of interest ($\lambda\to 2,3$) at
the end of the calculation. The desired solution will be a linear
combination of the above wave functions, \be\label{eqpsis} \Psi_\omega
(x) = A_+ (\Psi_0^+ + \Psi_1^+) + A_- (\Psi_0^- + \Psi_1^-)~. \ee
Asymptotically, it behaves as \be\label{eqpsias} \Psi_\omega(x) \approx
\left( A_+ +i e^{\frac{i\pi}{2\lambda}} \frac{\mathcal{A}
\Gamma(-1/\lambda)}{(2\omega)^{1-1/\lambda}} A_- \right)
e^{i\omega x} + \left( A_- - i
e^{-\frac{i\pi}{2\lambda}}\frac{\mathcal{A}
\Gamma(-1/\lambda)}{(2\omega)^{1-1/\lambda}} A_+\right)
e^{-i\omega x}~. \ee At large $r$ ($x\to +\infty$), the tortoise
coordinate~(\ref{eqtor}) may be approximated by \be\label{eqx0} x
\approx x_0 - \frac{1}{r} \ \ , \ \ \ \ x_0 = \int_0^\infty
\frac{dr}{f(r)} \ee and the potential~(\ref{eqpot}) is \be V(x)
\approx \xi^2 + \frac{1}{4}~. \ee Since we are interested in the
asymptotic form of quasi-normal frequencies, we may ignore the
potential ($V(x) \lesssim \omega^2$). We obtain the eigenfunction
for large $r$ ($x\sim x_0$) \be \Psi_\omega (x) \sim \sin \omega
(x-x_0)~, \ee where we applied the boundary condition $\Psi_\omega \to 0$
as $r\to\infty$ ($x\to x_0$).

This is matched by the linear combination~(\ref{eqpsias}) provided
\be\label{eqana} A_+ +i e^{\frac{i\pi}{2\lambda}}\frac{\mathcal{A}
\Gamma(-1/\lambda)}{(2\omega)^{1-1/\lambda}} A_- = -e^{-2i\omega
x_0} \left( A_- - i e^{-\frac{i\pi}{2\lambda}}\frac{\mathcal{A}
\Gamma(-1/\lambda)}{(2\omega)^{1-1/\lambda}} A_+ \right) \ee of
eigenfunction (eqs.~(\ref{eqpsi0}) and (\ref{eqpsi1})) in the
vicinity of the singularity.

Next, we approach the horizon ($x\to-\infty$) by analytically
continuing (\ref{eqpsis}) to negative $x$. This amounts to a
rotation by $-\pi$ in the complex $x$-plane. For $x<0$, we obtain
from~(\ref{eqpsi1}), using eqs.~(\ref{eqpot}) and (\ref{eqpsi0}),
\be\label{eqpsi1n} \Psi_1^\pm (x) = \frac{\mathcal{A}}{2i\omega}
e^{i\pi/\lambda}\int_0^{-x} \frac{dx'}{x^{\prime\, 1+1/\lambda}}
\left( e^{i(\omega +i\epsilon) (x+x')} - e^{-i(\omega -i\epsilon)
(x+x')} \right) e^{\mp i\omega x'} .\ee Taking the limit $x\to
-\infty$, we obtain the behavior near the horizon \be \Psi_\omega
(x) \approx \left( A_+ - i
e^{\frac{3i\pi}{2\lambda}}\frac{\mathcal{A}
\Gamma(-1/\lambda)}{(2\omega)^{1-1/\lambda}} A_- \right)
e^{i\omega x} + \left( A_- +i
e^{\frac{i\pi}{2\lambda}}\frac{\mathcal{A}
\Gamma(-1/\lambda)}{(2\omega)^{1-1/\lambda}} A_+\right)
e^{-i\omega x}~. \ee Since we want an ingoing wave
($\Psi_\omega\sim e^{i\omega x}$) at the horizon, we obtain the
constraint on the coefficients \be A_- +i
e^{\frac{i\pi}{2\lambda}}\frac{\mathcal{A}
\Gamma(-1/\lambda)}{(2\omega)^{1-1/\lambda}} A_+ = 0~. \ee For
compatibility with the other constraint~(\ref{eqana}), we ought to
have \be \left| \begin{array}{cc}  e^{-2i\omega x_0} +
e^{\frac{i\pi}{2\lambda}} \mathcal{C} & 1 \\ 1- \mathcal{C}
e^{-\frac{i\pi}{2\lambda}}e^{-2i\omega x_0} & \mathcal{C}
e^{-\frac{i\pi}{2\lambda}}\end{array} \right| = 0~, \ee where
$\mathcal{C} = i\mathcal{A}
\Gamma(-1/\lambda)(2\omega)^{-1+1/\lambda}$. We deduce \be
e^{2i\omega x_0} = 2e^{-\frac{i\pi}{2\lambda}}\mathcal{C}~,\ee
where we discarded terms which were of order higher than linear in
$\mathcal{C}$. Solving for $\omega$, we obtain the quasi-normal
frequencies \be \omega_n = \frac{n\pi}{x_0} + o(\ln n)~.
\label{OMX0} \ee Thus the asymptotic behaviour is completely
determined by the parameter $x_0$ (eq.(\ref{eqx0})).

In the case of MTZ black holes, $f(r)$ has four roots, \be r_\pm =
\frac{1}{2} \left( 1 \pm \sqrt{1+4G\mu} \right) \ \ , \ \ \ \ \bar
r_\pm = \frac{1}{2} \left( -1 \pm \sqrt{1-4G\mu} \right)~. \ee For
$G\mu > 0$, the horizon is at $r_+$ and $r_+ > 1$. We obtain \be
x_0 = \frac{1}{2(r_+-r_-)} \ln \frac{r_-}{r_+} + \frac{1}{2(\bar
r_+-\bar r_-)} \ln \frac{\bar r_-}{\bar r_+}~. \label{MTZEX} \ee
For large $G\mu$, the quasi-normal frequencies are \be \omega_n
\approx 2(1-i) r_+ n \label{MTZLM} \ee matching the behaviour in
{\em five}-dimensional AdS space.

For small (positive) $G\mu$, we find \be\  \omega_n \approx -2ni
\left( 1+ \frac{2G\mu}{\pi i} \ln G\mu \right) \label{MTZSM}~. \ee
Continuing to negative values of $G\mu$ is not straightforward.
The above discussion is not applicable, because going beyond the
horizon (as we approach the singularity) we encounter a potential
well at $r_- < r < r_+$ for $G\mu > -1/4$ (horizon $1/2 < r_+ <
1$) which admits bound states and alters the behaviour of
quasi-normal frequencies. The minimum of the potential provides a
lower bound to the frequencies. By setting $V'(r) = 0$, we find
the minimum at \be r_{min} = -2G\mu~. \ee At the minimum, \be
V(r_{min}) = - \left( \xi^2 + \frac{1}{4} \right) \left\{
\frac{1}{16(G\mu)^2} - 1 \right\}~. \ee It is also easily seen
that $V'' (r_{min}) > 0$, showing that $r=r_{min}$ is indeed a
minimum. The eigenfrequencies have imaginary part \be\label{eqlb}
\omega_I \ge -\sqrt{|V(r_{min})|} \label{bound}~, \ee which is
verified by numerical results. The lowest frequency is close to
the lower bound~(\ref{eqlb}).

In the TBH case, the horizon is given by \be r_+ = 2\Re (e^{i\pi
/6} s) \ \ , \ \ \ \ s = \left( \sqrt{\frac{1}{27} - (G\mu)^2} -
iG\mu \right)^{1/3}~, \ee where $|G\mu| < 3^{-3/2}$. The other two
roots of $f(r)$ are also real, \be r_- = -2 \Re (e^{-i\pi /6} s) \
\ , \ \ \ \ r_-' = 2\Im s~. \ee We obtain \be x_0 = -
\frac{r_-}{3r_-^2-1} \ln \frac{r_-}{r_+}- \frac{r_-'}{3r_-^{\prime
2}-1} \ln \frac{r_-'}{r_+}~. \label{TBHEX} \ee For small $G\mu$,
we have \be \omega_n \approx -2ni \left( 1 + \frac{4G\mu}{\pi i}
\ln G\mu \right) \label{TBHSM} \ee to be compared with the
behaviour of MTZ QNMs~(\ref{MTZSM}) near the transition point. The
discrepancy is of second order.

For large $G \mu$, the above formulae for the roots read  $$ r_+ =
\left( G\mu + \sqrt{(G\mu)^2 - \frac{1}{27}} \right)^{1/3}+ \left(
G\mu - \sqrt{(G\mu)^2 - \frac{1}{27}} \right)^{1/3} \approx \left(
2 G\mu \right)^{1/3}~, $$
$$ r_- = -e^{-\fr{i \pi}{3}} \left(
G\mu + \sqrt{(G\mu)^2 - \frac{1}{27}} \right)^{1/3}- e^{+\fr{i
\pi}{3}} \left( G\mu - \sqrt{(G\mu)^2 - \frac{1}{27}}
\right)^{1/3} \approx
 -e^{-\fr{i \pi}{3}} r_+~,
$$
\be r_-^\pr = (r_-)^* \approx -e^{+\fr{i \pi}{3}} r_+~. \ee Then \be \omega_n \approx n r_+ \left(\fr{3
\sqrt{3}}{4}-i \fr{9}{4}\right). \label{TBHLM} \ee
For $G\mu <0$, we encounter a potential well behind the horizon. Arguing as in the MTZ case, we obtain a lower bound
\be \omega_I \ge - \sqrt{|V(r_{min})|} \ee
In the TBH case, $r_{min} = -3G\mu$ and
\be\label{VminTBH} V(r_{min}) = -\left( \xi^2 + \frac{1}{4} \right) \left\{ \frac{1}{27(G\mu)^2} - 1 \right\} \ee

\section{Numerical Results}
\label{sec5}

\subsection{The Method}

Another method of studying the same problem is the procedure of
Horowitz and Hubeny \cite{HH}. We briefly review the method as we
applied it to our problem.
%We note that equations (\ref{schereq})
%may be written as $$f(r) \fr{d}{d r} \left(f(r) \fr{d}{d r}
%\Psi_\omega(r)\right) +\omega^2 \Psi_\omega(r) =
%\left(\xi^2+\fr{1}{4}\right)\fr{f(r)}{r^2} \Psi_\omega(r)~.$$ We
After performing the transformation $\Psi_\omega(r)=\psi_\omega(r) e^{-i
\omega r_*},$
the wave equation~(\ref{eqwav0}) becomes
\be f(r) \fr{d^2\psi_\omega(r)}{d
r^2}+ \left(\fr{d f(r)}{d r} -2 i \omega \right) \fr{d
\psi_\omega(r)}{d r} = \fr{\left(\xi^2+\fr{1}{4}\right)}{r^2}
\psi_\omega(r)~.\ee The change of variables $r=1/x$ (not to be confused with the tortoise coordinate~(\ref{eqtor}))
yields an equation of the form $$s(x) \left[(x-x_+)^2 \fr{d^2
\psi_\omega(x)}{d x^2}\right] +t(x) \left[(x-x_+) \fr{d
\psi_\omega(x)}{d x}\right]+ u(x)
\psi_\omega(x)=0~,$$ where $x_+ = 1/r_+$.
%$V(x) \equiv \left(\xi^2+\fr{1}{4}\right) x^2.$
It turns out that
$s(x), t(x)$ and $u(x)$ are polynomials of third degree for MTZ black holes
and of second degree for TBHs. Thus,
\begin{eqnarray} s(x) &=& s_0+s_1 (x-x_+)+s_2 (x-x_+)^2+s_3 (x-x_+)^3, \nonumber\\
t(x) &=& t_0+t_1
(x-x_+)+t_2 (x-x_+)^2+t_3 (x-x_+)^3,\nonumber\\
u(x) &=& u_0+u_1 (x-x_+)+u_2 (x-x_+)^2+u_3 (x-x_+)^3~.\nonumber\end{eqnarray}
Expanding the wavefunction
around the (inverse) horizon $x_+$,
\be\psi_\omega(x)=\sum_0^\infty a_n(\omega) (x-x_+)^n~,\ee
we arrive at a recurrence formula for the
coefficients,
\be a_n(\omega) = -\fr{1}{n(n-1)s_0+n
t_0+u_0}\sum_{m=n-3}^{n-1} [m (m-1) s_{n-m} +m t_{n-m} +u_{n-m}]
a_m(\omega)~.\ee We note that the few coefficients $a_m(\omega)$
with negative index $m$ which will appear for $n < 2$ should be
set to zero, while $a_0(\omega)$ is set to one. Since the wave function
should vanish at infinity $(r \rightarrow \infty, x=0),$ we deduce
%since the
%space is asymptotically AdS. We end up with the equation
\be
\psi_\omega(0) \equiv \sum_0^\infty a_n(\omega)
(-x_+)^n=0~.\label{HH} \ee
%Notice that the coefficients
%$a_n(\omega)$ are functions of $\omega,$ so equation (\ref{HH}) is
%actually an equation for $\omega.$
The solutions of this equation
are precisely the quasi-normal frequencies.

\subsection{MTZ Black Holes}

We compute the QNMs for the MTZ black holes by solving
eq.~(\ref{HH}) numerically. As a first step we draw the contours
$\Re[\psi_\omega(0)]=0$ and $\Im[\psi_\omega(0)]=0$ on the complex
$\omega-$plane and check the points of intersection. This provides
good initial values for the subsequent M\"{u}ller root-finding
technique \cite{NR}; in addition it provides an overview of the
(approximate) values of the quasi-normal frequencies. In
Fig.~\ref{MTZ5.0} we show sample contours for the case $r_+=5.0, \
\xi=1.0.$ We note that the parameter $\xi$ does not seem to play a
significant role in the behaviour of quasi-normal frequencies, so
we set it to a typical value $(\xi=1.0)$ from now on.

%%%%%%%%%%%%%%%%%%%%%%%%%%%%%%%%
\begin{figure}[ht]
\centering
\includegraphics[angle=-90,scale=0.6]{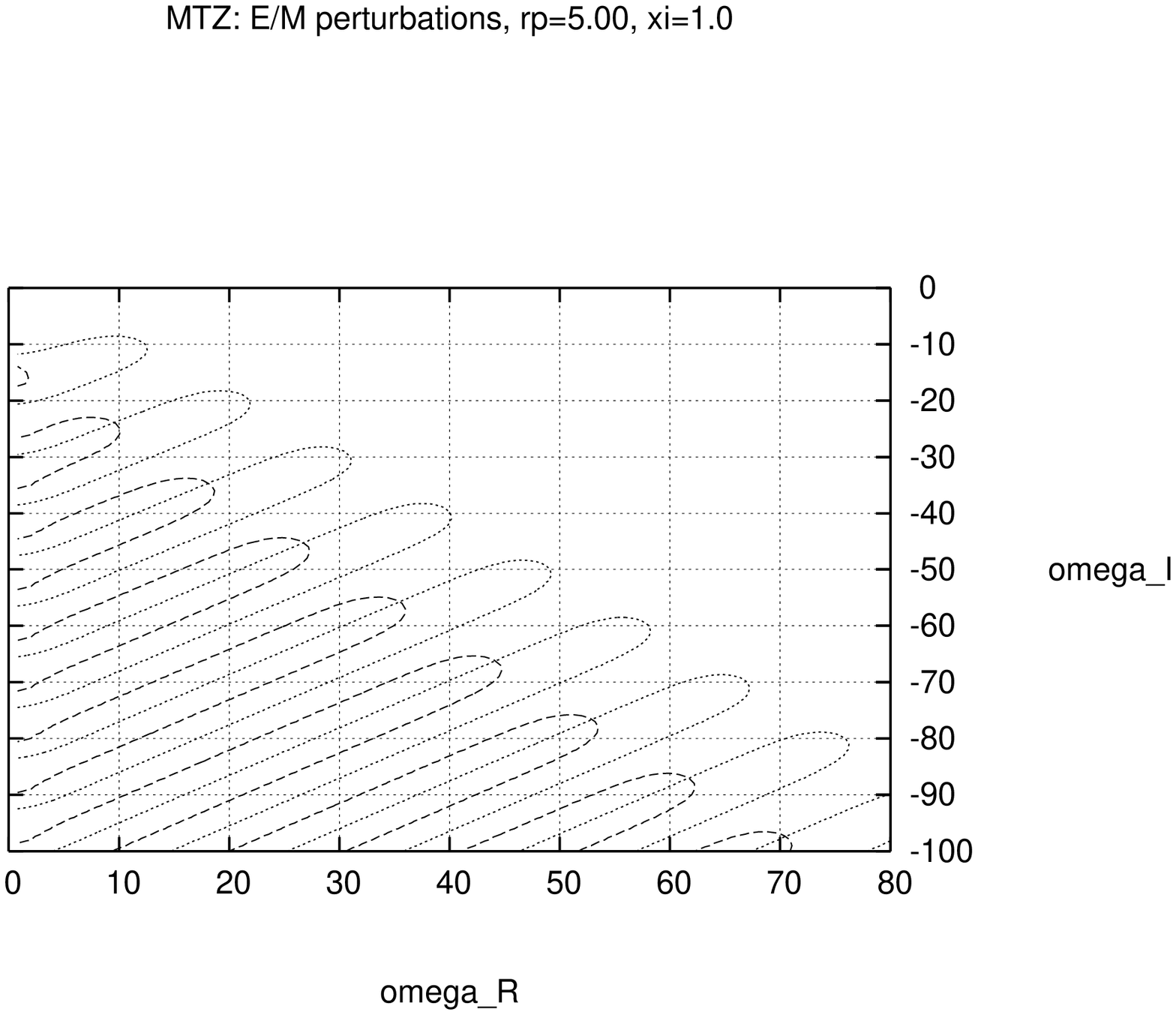}
\caption{Contours displaying the lines $\Re[\psi_0(\omega)]=0$
(dashed lines) and $\Im[\psi_0(\omega)]=0$} (dotted lines) on the
complex $\omega-$plane for MTZ Black Holes and $r_+=5.0.$
 \label{MTZ5.0}
\end{figure}
%%%%%%%%%%%%%%%%%%%%%%%%%%%%%%%%%%%%%%%%%%%%%%

From Fig.~\ref{MTZ5.0}, it is evident that the QNMs lie on a
straight line with {\it negative} slope and their spacing is more
or less constant. In fact the spacing changes a little as we move
to the right and eventually attains an asymptotic value, which
should be compared with the analytical results of section~\ref{sec4}.

Fig.~\ref{MTZ.97} refers to $r_+=0.97,$ a typical value of the
horizon radius smaller than the critical value $r_+=1.0.$ The
striking feature of the QNMs here is that the slope is {\it
positive.}  In addition, the intersections no longer lie along a
straight line and the spacing changes substantially as we move
through the QNMs.

%%%%%%%%%%%%%%%%%%%%%%%%%%%%%%%%
\begin{figure}[ht]
\centering
\includegraphics[angle=-90,scale=0.6]{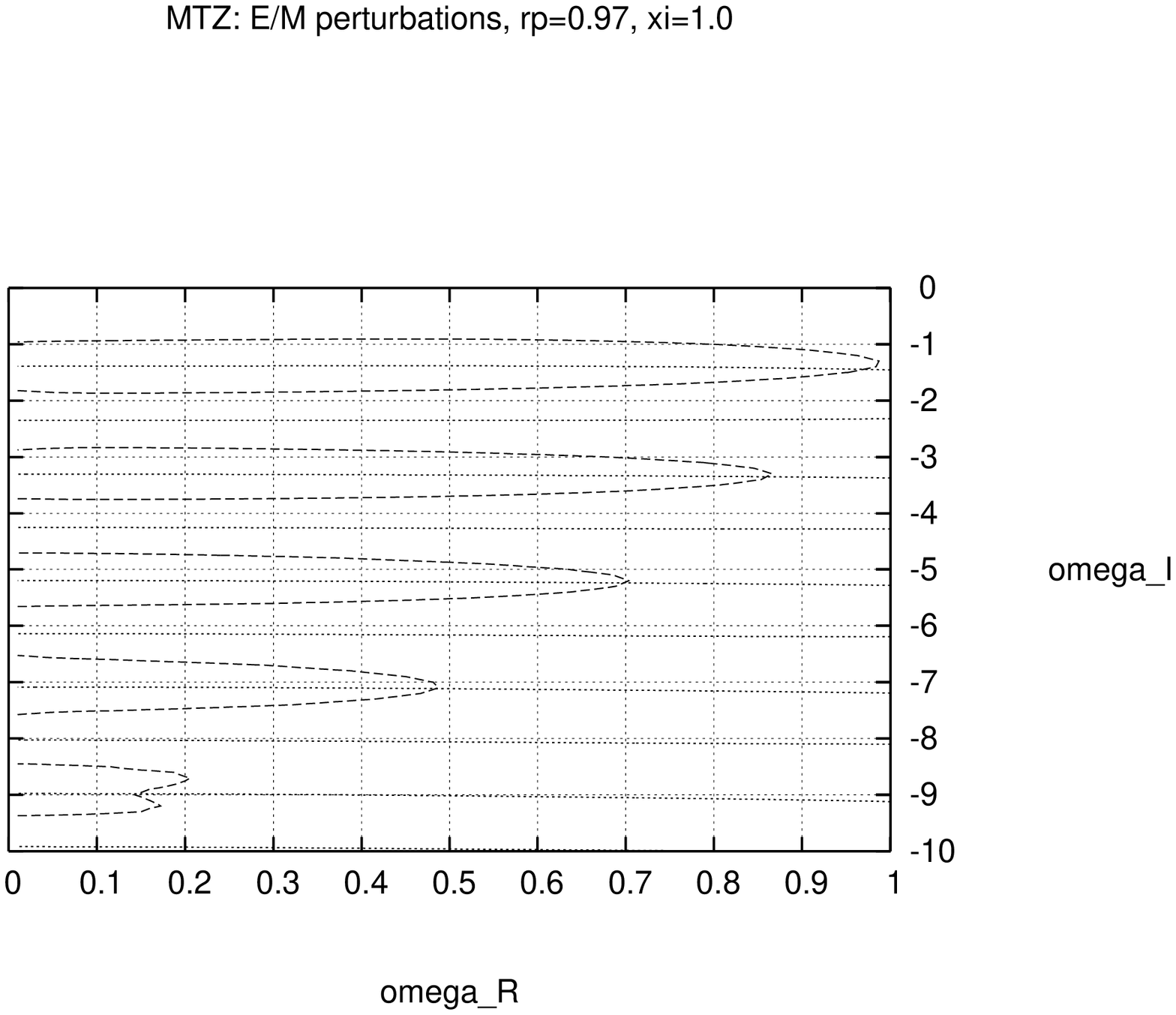}
\caption{Contours displaying the lines $\Re[\psi_0(\omega)]=0$
(dashed lines) and $\Im[\psi_0(\omega)]=0$} (dotted lines) on the
complex $\omega-$plane for MTZ Black Holes and $r_+=0.97.$
 \label{MTZ.97}
\end{figure}
%%%%%%%%%%%%%%%%%%%%%%%%%%%%%%%%%%%%%%%%%%%%%%

%We can compare the numerical results to the
To compare with analytical results, we note that
%analytical ones. We
%first recall the results of the analytic calculations. We remind
%the reader that $ G\mu=r^{2}_{+}-r_{+}.$ Equation
equations (\ref{OMX0}) and (\ref{MTZEX}) yield the asymptotic
expression for QNMs.
% in the limit of large horizon and mass.
%\be
%\Delta \omega_R = 2 r_+\,\,, \ \Delta \omega_I = -2 r_+~.
%\label{MLM} \ee where
%: $ \omega_n \approx 2(1-i) r_+ n, $ so
% the spacing
%$\Delta \omega \equiv \omega_{n+1} - \omega_n$ is the spacing of frequencies.
% We remark here that the assignment of numbers such
%as $m$ or $m+1$ above is conventional, since we may start counting
%from either the large or the small values of $\omega_R.$ The
%essential results are the absolute values of $\Delta \omega_R, \
%\Delta \omega_I$ and their relative sign.
We may check that the relative sign of $\Delta\omega_R$ and
$\Delta\omega_I$ is negative, therefore a larger $\omega_R$ will
correspond to an algebraically smaller $\omega_I,$ as shown in
Fig.~\ref{MTZ5.0}.
%In this limit the spacing is proportional to$r_+.$
% so it makes sense to divide our results by $r_+.$

For small mass ($r_+ \approx 1$), the analytical result~(\ref{MTZSM})
 yields the asymptotic expression for the spacing of QNMs,
%approximate
%expression for the QNMs in the limit of small horizon and mass,
%$ \omega(n) \approx -2ni \left( 1 + \frac{2 G\mu}{\pi i} \ln G\mu
%\right)$, which yields for the spacings:
\be \Delta \omega_R
\approx -\fr{4 G \mu}{\pi}\ln G \mu, \ \Delta \omega_I \approx
-2~. \label{MSM} \ee
%In this case division by $r_+$ does not help.

In Table~\ref{TMTZL}, we show the results for some indicative
values of the horizon radius with $r_+ \ge 1.0.$ We observe that
the agreement between analytical and numerical results is
quite good. The agreement is exact for $r_+=1.0.$ In fact, the
slope tends to minus infinity as we approach $r_+=1.0.$ We should point out
that $\omega_I<0,$ always, so there is no sign of instability.

\begin{table}[ht]
\begin{center}
\begin{tabular}{|c|c|c|c|c|c|c|}

\hline

$r_+$ &  $G \mu$ & $T$ & $\Delta \omega_R^{Ana}$ & $\Delta
\omega_I^{Ana}$ & $\Delta \omega_R^{Num}$ & $\Delta \omega_I^{Num}$
\\ \hline

$10.000$ & $90.00$ & $3.023$ &   $18.96$  & $-20.26$ & $ 19.00 $ &
$-20.15$ \\ \hline

$5.000$ & $20.00$ & $1.434$ &   $8.91$  & $-10.25$ & $ 8.77 $ &
$-10.45$ \\ \hline

$2.000$ & $2.000$ & $0.477$ &   $2.75$  & $-4.21$ & $ 2.70 $ & $
-4.14 $ \\ \hline

$1.050$ & $0.0525$ & $0.175$ &   $0.32$  & $-2.15$ & $ 0.27 $ &
$-2.16$ \\ \hline

$1.000$ & $0.000$ & $0.160$ &   $0.00$  & $-2.00$ & $ 0.00 $ &
$-2.00$ \\ \hline

\end{tabular}
\end{center}
\caption{Comparison analytical and numerical values of QNMs of EM
perturbations of MTZ Black Holes.} \label{TMTZL}
\end{table}
%##############################################################

In Table~\ref{TMTZS} we show the QNMs for $r_+=0.97,$ as an example of horizon radius
smaller than
the critical value $r_+=1.0$, along with the differences between consecutive
values of QNMs. Note that in this case $G \mu = -0.029, \ T=0.150.$
\begin{table}[ht]
\begin{center}
\begin{tabular}{|c|c|c|c|}

\hline

$\omega_R^{Num}$ & $\omega_I^{Num}$ & $\Delta \omega_R^{Num}$ &
$\Delta \omega_I^{Num}$  \\ \hline

$0.973$  & $-1.496$ &    -      &     -     \\ \hline

$0.864$  & $-3.351$ & $0.109$  & $1.905$  \\ \hline

$0.701$  & $-5.239$ & $0.163$  & $1.888$  \\ \hline

$0.486$  & $-7.114$ & $0.215$  & $1.875$  \\ \hline

$0.143$  & $-8.980$ & $0.343$  & $1.866$  \\ \hline

\end{tabular}
\end{center}
\caption{Numerical results for quasi-normal frequencies of EM perturbations for
MTZ Black Holes with $r_+=0.97.$ In the last two columns we list
the differences of consecutive QNMs.} \label{TMTZS}
\end{table}
%##############################################################
These are the exact QNMs presented in Fig.~\ref{MTZ.97}. It
appears that, apart from the change of the sign of the slope,
there is a novel phenomenon: the quasi-normal frequencies converge
toward the imaginary axis, i.e., their real part decreases. There
is only a {\em finite} number of QNMs for $r_+ <1.0.$ This
behaviour has already been predicted analytically in
section~\ref{sec4}; eq.~(\ref{bound}) yields a prediction for the
lowest possible imaginary part of the frequencies. The result is
$\omega_I \ge -9.54 i,$ which is indeed respected by the imaginary
parts appearing in Table \ref{TMTZS}. Although not fully
justified, using the analytical result~(\ref{MSM}), we deduce the
estimates $|\Delta \omega_R^{Ana}| \approx 0.131, \ \ \ |\Delta
\omega_I^{Ana}| \approx 2.000.$ Checking against the numerical
values in Table~\ref{TMTZS}, we see that the values in the third
column (real part) are of the same order of magnitude as the
analytical prediction, whereas the values in the fourth column
(imaginary part) are quite close to the analytical estimate $2.$
However, the variations are significant. If one further decreases
the value of the horizon, the number of QNMs decreases until they
finally disappear. The behavior of QNMs for such horizons may be
viewed as modifications of the critical point $r_+=1.0$. In the
latter case, the first quasi-normal frequency is $\omega = 1.0-1.5
i$, followed by frequencies with imaginary parts $\omega_I = -3.5,
-5.5, -7.5,\dots $. Unlike the points below criticality, for
$r_+=1.0,$ the real parts do not change and there is no lower
bound.

To summarize our findings, for $r_{+} > 1$, Fig.~\ref{MTZ5.0}
shows that the QNMs have a negative slope and hence large values
of $\omega_{R}$ correspond to large values of $\omega_{I}$. Then
as $r_{+}\rightarrow 1$ the slope tends to minus infinity. At
$r_{+} = 1$, $\omega_{R}=1,$ while the spacing of $\omega_{I}$ is
exactly $2.$ Note that the critical point $r_{+} = 1$ corresponds
to the pure AdS configuration of (\ref{muzero}). For $r_{+} < 1$,
Fig.~\ref{MTZ.97} shows that the QNMs have a positive slope
and hence large values of $\omega_{R}$ correspond to small values
of $\omega_{I}$. We attribute this behavior to a {\it phase
transition,} as one passes through the critical value $r_+=1,$ of
the TBH to the MTZ black hole configuration.

Notice that for any perturbation of a black hole background there
are two characteristic parameters that control its behaviour: the
oscillation time scale which is given by $\tau_R \equiv
1/\omega_{R}$ and the damping time scale given by $\tau_I \equiv
1/\omega_{I}$. In the case of $r_{+}
> 1$ the scalar field is absorbed by the black hole and the damping
time scale $\tau_{I}$ is small, so the perturbations in this case
fall off rather rapidly with time. For $r_{+} < 1$ the black hole
is dressed up with the scalar field and the damping time scale
$\tau_{I}$ is large and perturbations last longer. At the critical
point of $r_+=1$ we have a change of slope, indicating a transient
configuration. This behaviour may be associated with the second
order phase transition discussed in section~\ref{sec3}.

\subsection{Topological Black Holes}

We now turn to the case of topological black holes (TBHs). The asymptotic
value of quasi-normal frequencies for $r_+\ge 1$ is given by
%We first recall the
%results of the analytic calculations. The quantity $G \mu$ is
%given by: $ G\mu=(r^{3}_{+}-r_{+})/{2}.$ As in the previous case,
eqs~(\ref{OMX0}) and (\ref{TBHEX}).
%$ \omega_n = n r_+ \left(\fr{3 \sqrt{3}}{4}-i
%\fr{9}{4}\right),$ so the spacing $\Delta \omega \equiv
%\omega_{m+1} - \omega_m$ yields: \be \Delta \omega_R = \fr{3
%\sqrt{3}}{4} r_+ \approx 1.30 r_+\,\,, \ \Delta \omega_I =
%-\fr{9}{4} r_+ \approx -2.25 r_+~.\ee We remark here that the
%assignment of numbers such as $m$ or $m+1$ above is conventional,
%since we may start counting from either the large or the small
%values of $\omega_R.$ The real results are the absolute values of
%$\Delta \omega_R, \ \Delta \omega_I$ and their relative sign.
The relative sign of $\Delta \omega_R,$ and $\Delta \omega_I$ is
again negative.
%We observe that in this limit the spacing is
%proportional to $r_+,$ so it makes sense to divide our results by
%$r_+.$

At small masses, eq.~(\ref{TBHSM}) provides an approximate
expression for QNMs in the limit of small horizon (mass)
%$ \omega_n \approx -2ni \left( 1+ \frac{4 G \mu}{\pi i} \ln G\mu \right),$
which yields the spacings \be \Delta \omega_R
\approx -\fr{8 G \mu}{\pi}\ln G \mu\,\,, \ \Delta \omega_I \approx
-2~. \label{TSM} \ee
The comparison of these analytical results to the numerical results of section~\ref{sec5}
%for the TBH can be compared with the
%numerical calculation of the QNMs for the EM perturbations. The
%results are
is shown in Table \ref{TTBHL}. We have used
values for the horizon which correspond, through eq.~(\ref{rhor}),
to the respective values
of MTZ black holes that we showed above in Table~\ref{TMTZL} (matching corresponding temperatures).
%One may check
%that the temperatures appearing in Table \ref{TTBHL} equal the
%ones appearing in Table \ref{TMTZL}.
The agreement
between analytical and numerical results is very good.
%, similar remarks apply as
%in the MTZ case.
In addition, $\omega_I$ is again negative,
showing no sign of instability.

\begin{table}[ht]
\begin{center}
\begin{tabular}{|c|c|c|c|c|c|c|}

\hline

$r_+$ &  $G \mu$ & $T$ & $\Delta \omega_R^{Ana}$ & $\Delta
\omega_I^{Ana}$ & $\Delta \omega_R^{Num}$ & $\Delta \omega_I^{Num}$
\\ \hline

$12.692$ & $1015.43$ & $3.023$ &   $16.43$  & $-28.55$ & $ 16.50 $
& $-28.50$ \\ \hline

$6.055$ & $107.97$ & $1.434$ &   $7.73$  & $-13.60$ & $ 7.65 $ &
$-13.70$ \\ \hline

$2.155$ & $3.93$ & $0.477$ &   $2.41$  & $-4.78$ & $ 2.52 $ &
$-4.81$\\ \hline

$1.050$ & $0.054$ & $0.175$ &   $0.29$  & $-2.16$ & $ 0.26 $ &
$-2.16$ \\ \hline

$1.000$ & $0.000$ & $0.160$ &   $0.00$  & $-2.00$ & $ 0.00 $ &
$-2.00$ \\ \hline

\end{tabular}
\end{center}
\caption{Comparison of analytical and numerical values of QNMs of EM
perturbations for TBHs.} \label{TTBHL}
\end{table}

Next we show the QNMs for $r_+=0.97$ (eq.~(\ref{rhor}) yields
approximately the same value as in the MTZ case for the same temperature), along with
differences between consecutive values of quasi-normal frequencies.
Note that in this case, $G \mu =
-0.029, \ T=0.150.$

\begin{table}[ht]
\begin{center}
\begin{tabular}{|c|c|c|c|}

\hline

$\omega_R^{Num}$ & $\omega_I^{Num}$ & $\Delta \omega_R^{Num}$ &
$\Delta \omega_I^{Num}$  \\ \hline

$0.972$  & $-1.446$ &    -      &     -     \\ \hline

$0.859$  & $-3.352$ & $0.113$  & $1.906$  \\ \hline

$0.689$  & $-5.240$ & $0.170$  & $1.888$  \\ \hline

$0.453$  & $-7.115$ & $0.236$  & $1.875$  \\ \hline

\end{tabular}
\end{center}
\caption{Numerical results for quasi-normal frequencies of EM perturbations for
TBHs with $r_+=0.97.$ In the last two columns
we list the differences of consecutive QNMs.} \label{TTBHS}
\end{table}
%##############################################################

Similar behaviour to MTZ black holes is observed: the slope below
criticality becomes {\em positive}, while for $r_+>1$ it is {\em
negative.} Again, to compare with analytical expressions,
eq.~(\ref{TSM}) gives the estimate $|\Delta \omega_R^{Ana}|
\approx 0.259, \ \ \ |\Delta \omega_I^{Ana}| \approx 2.000.$
Checking against the entries in Table~\ref{TTBHS}, we find that
the values in its third and fourth columns (real and imaginary
parts, respectively) are of the same order of magnitude as the
analytical estimates, albeit with sizeable variations. The lower
bound for the imaginary part, given by eqs.~(\ref{bound}) and
(\ref{VminTBH}), reads in this case,
 $\omega_I \ge -7.33,$ which is close to the lowest
imaginary part in Table~\ref{TTBHS}.

\section{Conclusions}
\label{sec6}

We calculated the QNMs of electromagnetic perturbations of the MTZ
black hole and topological black holes. We performed this
calculation both analytically and numerically with fairly good
agreement. We found that there is a change in the slope of the
QNMs as we decrease the value of the horizon radius below a
critical value, and we argued that this change signals a phase
transition of a vacuum topological black hole toward the MTZ
black hole with scalar hair.

One may attribute this change in behaviour to the dynamics of the
scalar field and associate it with a second-order phase transition
\cite{Martinez:2004nb}. For small black holes ($r_{+}<1$) the
scalar field is dressing up the bare topological black hole
introducing an order parameter $\lambda_{\phi}$ (see
(\ref{difference1})) which controls the dynamics of the scalar
field.
 A second-order phase transition for small black holes
occurs in more general configurations (including charge, etc) as
studies of general scalar, electromagnetic and gravitational
perturbations show~\cite{future}. We have also found that the
vacuum AdS solution at the critical temperature is a transient
configuration of the change of phase of the topological black hole
to a  configuration with scalar hair.
%In all cases we found a second order
%phase transition for small black holes.

One interesting aspect of small MTZ black holes is that the
quasi-normal frequencies converge toward the imaginary axis, i.e.,
their real part decreases and after the first few quasi-normal
frequencies, it vanishes, indicating that for $r_+ <1$ there are
only a finite number of QNMs. We showed that the finite number of
such modes for small horizons ($r_+ <1$) is due to the existence of
bound states behind the horizon, which is an unobservable region.
This curious phenomenon is worthy of further investigation.

\section*{Acknowledgements}

We thank Kostas Kokkotas for his valuable assistance in the
numerical applications of this work. The work of S.~M.~was
partially supported by the Thailand Research Fund. G.~K. and E.~P.
 were supported by (EPEAEK II)-Pythagoras (co-funded by the
European Social Fund and National Resources). G.~S.~was supported
in part by the US Department of Energy under grant
DE-FG05-91ER40627.

\end{document}